\documentclass[twocolumn]{jpsj2} 

\def\runtitle{Instructions for the Preparation of Manuscript}
\def\runauthor{Online-Journal Subcommittee}

\title{
$^{59}$Co Nuclear Quadrupole Resonance Studies of Superconducting and Non-superconducting $Bi$layer Water Intercalated Sodium Cobalt Oxides Na$_{x}\mathrm{CoO_{2}} \cdot$$y\mathrm{H_{2}O}$
}
\author{
C. Michioka, H. Ohta, Y. Itoh, and K. Yoshimura
}

\inst{
Department of Chemistry, Graduate School of Science,\\
Kyoto University, Kyoto 606-8502, Japan \\ 
}

\recdate{\today}

\abst{
We report $^{59}$Co nuclear quadrupole resonance (NQR) studies of {\it bi}layer water intercalated sodium cobalt oxides Na$_{x}\mathrm{CoO_{2}} \cdot$$y\mathrm{H_{2}O}$ (BLH) with the superconducting transition temperatures, 2 K $<$ $T\mathrm{_{c}}$ $\le$ 4.6 K, as well as a magnetic BLH sample without superconductivity.  We obtained a magnetic phase diagram of $T\mathrm{_{c}}$ and the magnetic ordering temperature $T\mathrm{_{M}}$ against the peak frequency $\nu_{3}$ of the $^{59}$Co NQR transition $I_z = \pm 5/2 \leftrightarrow \pm 7/2$ and found a dome shape superconducting phase. The $^{59}$Co NQR spectrum of the non-superconducting BLH shows a broadening below $T\mathrm{_{M}}$ without the critical divergence of 1/$T_{1}$ and 1/$T_{2}$, suggesting an unconventional magnetic ordering. The degree of the enhancement of $1/T_{1}T$ at low temperatures increases with the increase of $\nu_{3}$ though the optimal $\nu_{3} \sim$ 12.30 MHz. In the Na$_{x}\mathrm{CoO_{2}} \cdot$$y\mathrm{H_{2}O}$ system, the optimal-$T\mathrm{_{c}}$ superconductivity emerges close to the magnetic instability. $T\mathrm{_{c}}$ is suppressed near the phase boundary at $\nu_{3} \sim$ 12.50 MHz, which is not a conventional magnetic quantum critical point.
}

\kword{
superconductivity,sodium cobalt oxide,nuclear quadrupole resonance
}

\begin{document}
\maketitle

The discovery of superconductivity of $bi$layer water intercalated sodium cobalt oxides Na$_{x}\mathrm{CoO_{2}} \cdot$$y\mathrm{H_{2}O}$ (BLH) has renewed our interests in itinerant triangular lattice systems. \cite{Takada} After much experimental efforts have been devoted to find the chemical parameters on the superconducting transiton temperature $T\mathrm{_{c}}$, the complicated and delicate structural parameters on the Na vacancies, the intercalated water molecules and oxonium ions, which partially occupy the sodium site, \cite{Takada2} have been revealed. However, the microscopic roles of these parameters in characterizing the superconductors are still poorly understood. 

Recently we successfully synthesized powder samples of Na$_{x}\mathrm{CoO_{2}} \cdot$$y\mathrm{H_{2}O}$ with different superconducting transition temperature $T\mathrm{_{c}}$ by a soft chemical method by utilizing so-called duration effect. \cite{Ohta} The $T\mathrm{_{c}}$ of a series of samples shows a systematic change with the increase of time duration after synthesis of $bi$layer Na$_{x}\mathrm{CoO_{2}} \cdot$$y\mathrm{H_{2}O}$ being kept in a humidified chamber (75 $\%$ relative humidity atmosphere). The another conditions of humidity and temperature were independently found to lead to different duration effects, where $T\mathrm{_{c}}$ also shows a systematic change. \cite{Bernes} In contrast to the X-ray and neutron diffraction techniques, the $^{59}$Co nuclear quadrupole resonance (NQR) is a powerful technique to investigate local electronic state around cobalt atoms. Thus we have performed $^{59}$Co nuclear quadrupole resonance (NQR) studies on $bi$layer Na$_{x}\mathrm{CoO_{2}} \cdot$$y\mathrm{H_{2}O}$ prepared by using the duration effect to characterize the electronic states of CoO$_2$ planes and the electron spin dynamics microscopically. 

The powder samples of $bi$layer water intercalated sodium cobalt oxides were prepared by the same way as mentioned in ref. 3. A $mono$layer water intercalated sodium cobalt oxide (MLH) was synthesized by a deintercalation method of water molecules from BLH in a dry N$_2$ gas flowing atmosphere.

\begin{figure}
\begin{center}
\includegraphics[width=0.65\linewidth]{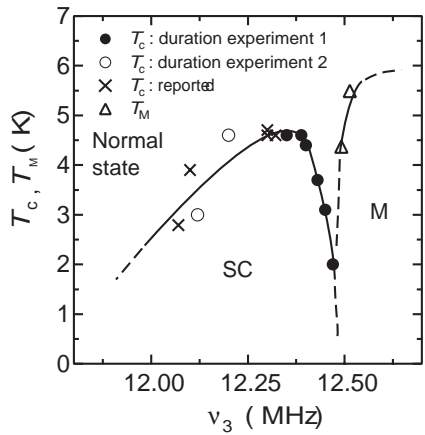}
\end{center}
\caption{\label{fig:phase}
Magnetic phase diagram of BLH Na$_{x}\mathrm{CoO_{2}} \cdot$$y\mathrm{H_{2}O}$ decided by NQR measurements. $T\mathrm{_{c}}$ and $T\mathrm{_{M}}$ are plotted against $\nu_{3}$ at 10 K. Filled and open circles represent the data for the samples synthesized by duration effect from two different initial conditions of the parent compound. Cross marks are the data reported in ref. 7. Triangles are the magnetic ordering temperature $T\mathrm{_{M}}$ decided from the temperature dependence of the nuclear spin-lattice relaxation rate (1/$T_1$). SC and M denote superconducting and magnetic states, respectively.
}
\end{figure}

The $^{59}$Co (nuclear spin $I$=7/2) NQR experiments have been done for the powder samples by utilizing a coherent-type pulsed NMR spectrometer. The $^{59}$Co nuclear spin-lattice relaxation time $^{59}T_{1}$ was measured by an inversion recovery technique. The spin-echo signal $M(t)$ was measured as a function of long delay time $t$ after an inversion pulse, and $M(\infty)[\equiv M(t>10T_1)]$ was recorded. All the recovery corves were measured at the peak freqency of $\nu_{3}$, which corresponds to the the transition $I_z = \pm 5/2 \leftrightarrow \pm 7/2$.
In order to estimate $^{59}T_{1}$, the recovery curves $p(t)\equiv 1-M(t)/M(\infty)$ were fitted by a theoretical
$p(t)=p(0)[(3/7)\mathrm{e}^{-3t/{T_1}}+(100/77)\mathrm{e}^{-10t/{T_1}}+(3/11)\mathrm{e}^{-21t/{T_1}}]$
for the transition ($I_z = \pm 5/2 \leftrightarrow \pm 7/2$),
where $p(0)$ and $T_1$ are fit parameters. \cite{Chepin} Agreement of the theoretical and experimental recovery curves was totally satisfactory except for magnetically ordered states and also except for the samples near the boundary between the superconducting and magnetic phases in Fig. \ref{fig:phase}. 
The temperature dependence of the transverse relaxation curves was also measured with using a spin echo technique for non-superconducting BLH.
A production of Lorentzian ($T_{2\mathrm{L}}$) and Gaussian ($T_{2\mathrm{G}}$) terms with double time constants, $m(2\tau) = m(0) \mathrm{exp}[-(2\tau/T_{2\mathrm{L}})-1/2(2\tau/T_{2\mathrm{G}})^2]$, was fitted to the data.

\begin{figure}
\begin{center}
\includegraphics[width=0.88\linewidth]{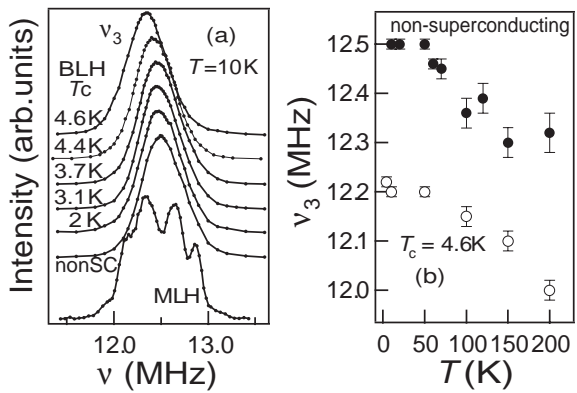}
\end{center}
\caption{\label{fig:spe_all}
(a) $^{59}$Co NQR spectra of $\nu_{3}$ in superconducting BLH Na$_{x}\mathrm{CoO_{2}} \cdot$$y\mathrm{H_{2}O}$ with $T\mathrm{_{c}}$ = 2, 3.1, 3.7, 4.4, and 4.6 K, non-superconducting BLH and MLH measured at 10 K. These resonance lines correspond to the transition ($I_z = \pm 5/2 \leftrightarrow \pm 7/2$). (b) Temperature dependence of $\nu_{3}$ in BLH with non-superconducting (right side in the dome shape superconducting phase in Fig. \ref{fig:phase}) and $T\mathrm{_{c}}$ = 4.6 K (left side in the dome shape superconducting phase in Fig. \ref{fig:phase}) samples.}
\end{figure}

Figure \ref{fig:spe_all} (a) shows the NQR spectra at 10 K on superconducting ($T\mathrm{_{c}}$ = 2, 3.1, 3.7, 4.4, and 4.6 K) and non-superconducting BLH as well as MLH Na$_{x}\mathrm{CoO_{2}} \cdot$$y\mathrm{H_{2}O}$. These resonance lines arise from the transition, $I_z = \pm 5/2 \leftrightarrow \pm 7/2$. In the superconductors BLH, the peak freqency of $\nu_{3}$ decreases as temperature increases as shown in Fig. \ref{fig:spe_all} (b).

The MLH compound has an alternating stacking structure of Na/H$_2$O and CoO$_2$ layers. Four peaks in the NQR spectrum corresponding to $\nu_{3}$ suggest that at least four cobalt sites exist in the MLH. But we cannot assign these sites because of lack of information on the Na/H$_2$O local structure. The sodium ions and water molecules occupy just upper sites of oxygen sites along the $c$-axis in the MLH. \cite{Takada2} The more than four cobalt sites with different NQR frequency $\nu_{\mathrm{Q}}$ may be related with Na vacancy superlattice and oxonium ordering. 

In the case of BLH, only a broad single peak of $\nu_{3}$ appears in the NQR spectrum. The intercalated water molecules smear out the local electric field gradient around several cobalt sites. The BLH compounds, whose NQR spectra are shown in Fig. \ref{fig:spe_all}, were synthesized by a single set of time duration treatment mentioned in ref. 3. As the time duration in the humidified chamber and $T\mathrm{_{c}}$ increase, the peak frequency $\nu_{3}$ of the NQR spectrum decreases without the change of line shape. If the oxygen content changes with the time duration, the NQR spectrum would be expected to show asymmetrically broadening as in the case of high-$T\mathrm{_{c}}$ cupper oxides. Thus the oxygen content in BLH is thought to be stoichiometric. From the analysis of X-ray powder diffractions, \cite{Ohta} the lattice constants were nearly the same for the present samples. The cobalt valence in the BLH system is reported to be +3.4 and does not change so much. \cite{Sakurai} Thus, not the doped carriers but the local crystal shrink of CoO$_{6}$ octahedrons changes with the time duration accompanied by the decrease of $\nu_{3}$. The distortion of CoO$_{6}$ octahedron may cause some change in the electronic state of CoO$_{2}$ planes and then increase $T\mathrm{_{c}}$. Therefore, the value of $\nu_{3}$ can be a measure of the change in the electronic state.

Figure \ref{fig:phase} is a magnetic phase diagram of $T\mathrm{_{c}}$ plotted against $\nu_{3}$ at 10 K, which was drawn from the results by the magnetic susceptibility and $\nu_{3}$ from the $^{59}$Co NQR spectra measurements. $T\mathrm{_{M}}$ was determined by measurements of the nuclear spin relaxation rates as shown later. The closed circles in Fig. \ref{fig:phase} were obtained for the present BLH samples in Fig. \ref{fig:spe_all} (a). The open circles were obtained for the other BLH samples with the different duration effect using different initial mother compounds. The crosses in Fig. \ref{fig:phase} denote the previous data on the accidentally obtained samples with various $T\mathrm{_{c}}$'s. \cite{Ihara} Because we study the samples with systematically changing $T\mathrm{_{c}}$ in this work, the consistency with the previous work suggests that the samples in ref. 7 are involved in the series of our samples. We found that $T\mathrm{_{c}}$ is suppressed near the boundary between the superconducting and the magnetic phases. The optimal $T\mathrm{_{c}}$ and the highest suprconducting volume fraction are realized in the sample with $\nu_3 \sim 12.30$ at 10 K. 

The domed superconducting phase is similar to those of the high $T\mathrm{_{c}}$ cuprates. The similar dome shape phase diagram of $T\mathrm{_{c}}$ as a function of the sodium content in Na$_{x}\mathrm{CoO_{2}} \cdot$$y\mathrm{H_{2}O}$ system is reported in ref. 8. However, the formal cobalt valence does not change with the different ($\nu_{3}$, $T\mathrm{_{c}}$) samples of the same sodium content. If the sodium content can be changed from a sample to another, the additional oxonium ions should compensate for the loss of charge to make cobalt valence almost invariant. \cite{Sakurai} Therefore, we believe that the coordinate of ($\nu_{3}$, $T\mathrm{_{c}}$) specifies the degree of the distortion of CoO$_6$ octahedron, which may change the band structure.

\begin{figure}
\begin{center}
\includegraphics[width=0.68\linewidth]{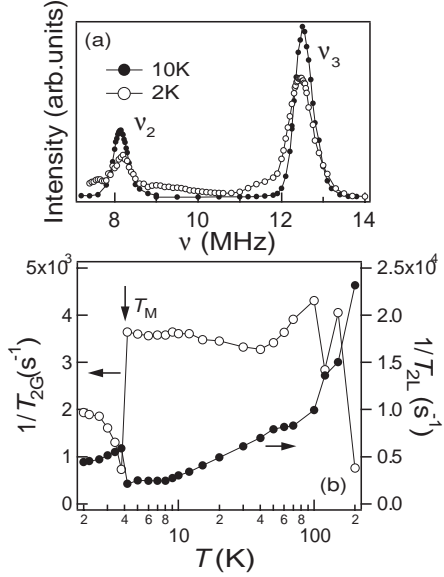}
\end{center}
\caption{\label{fig:spe_nonsuper}
(a) NQR spectra of non-superconducting BLH measured at 2 and 10 K. (b) Temperature dependence of 1/$T_{2\mathrm{G}}$ (open circles) and 1/$T_{2\mathrm{L}}$ (closed circles) in non-superconducting BLH measured at the peak frequency $\nu_{3}$ (12.50 MHz at 10 K). $T\mathrm{_{M}}$ is a magnetic ordering temperature.
}
\end{figure}

We obtained several non-superconducting BLH samples by using the different initial and humidity conditions in the synthesis process. Figure \ref{fig:spe_nonsuper} (a) shows the $^{59}$Co NQR spectra in one of the non-superconducting BLH samples at 2 and 10 K. The separable peaks at about 8.15 and 12.50 MHz correspond to the transition lines $\nu_{2}$ ($I_z = \pm 3/2 \leftrightarrow \pm 5/2$) and $\nu_{3}$ ($I_z = \pm 5/2 \leftrightarrow \pm 7/2$), respectively. At 10 K, the line shape and width are similar to those of the superconducting BLH samples except for a slight difference in the peak position of the resonance line. The line shapes of the superconducting BLH samples do not change so much in the superconducting state at 2 K. However, the line shape of the non-superconducting sample drastically changes at 2 K; the $\nu_{2}$ line broadens both toward the lower and upper frequency sides while the $\nu_{3}$ line shows broadening only to the lower frequency side. If these changes in the NQR spectra arise from changes in electric field gradients with cooling, the change in the $\nu_{3}$ line should be scaled by that in the $\nu_{2}$ line. But the actual change was lager in $\nu_{2}$ than that in $\nu_{3}$. Therefore, the asymmetric broadenings in the $\nu_{2}$ and $\nu_{3}$ lines indicate the occurrence of an internal magnetic field. In general, the NQR spectrum of $I = 7/2$ would show broadening at the upper frequency side of $\nu_{3}$ under the small internal magnetic field parallel to the principal axis of the electric field gradient tensor ($c$-axis), \cite{Nishihara} suggesting that the small internal magnetic field should point in the CoO$_{2}$ plane in the present case. The broad but not split $\nu_{2}$ line suggests that the internal magnetic field distributes in the plane. The presence of such a small internal magnetic field in non-superconducting BLH is similar to that previously reported. \cite{Ihara}

Figure \ref{fig:spe_nonsuper} (b) shows the temperature dependence of the relaxation rates 1/$T_{2\mathrm{L}}$ and 1/$T\mathrm{_{2G}}$ in the non-superconducting BLH sample. The primary contribution to 1/$T_{2\mathrm{L}}$ is a $T_{1}$ process. 1/$T\mathrm{_{2G}}$ is a nuclear spin-spin relaxation rate. 1/$T\mathrm{_{2G}}$ increases with cooling below 50 K, indicating that an indirect nuclear spin-spin coupling enhances and the coupling constant with the static spin susceptibility increases below 50 K. 
At high temperatures above 100 K, the narrowing effect such as sodium motion may increase 1/$T_{2\mathrm{L}}$ and decrease 1/$T_{2\mathrm{G}}$. At 4.2 K, both 1/$T_{2\mathrm{L}}$ and 1/$T\mathrm{_{2G}}$ show abrupt changes. Taking into account the existence of the static internal magnetic field, the abrupt change in 1/$T_{2\mathrm{L, G}}$ and an anomaly in 1/$T_1$ (shown later), a kind of static magnetic ordering occurs at $T\mathrm{_{M}} \sim$ 4.2 K. The distributed internal magnetic field and the absence of conventional critical slowing down effect on $T_{2}$ and $T_{1}$ indicate an unconventional magnetic ordering such as a quasi-long range ordering.

\begin{figure}
\begin{center}
\includegraphics[width=0.75\linewidth]{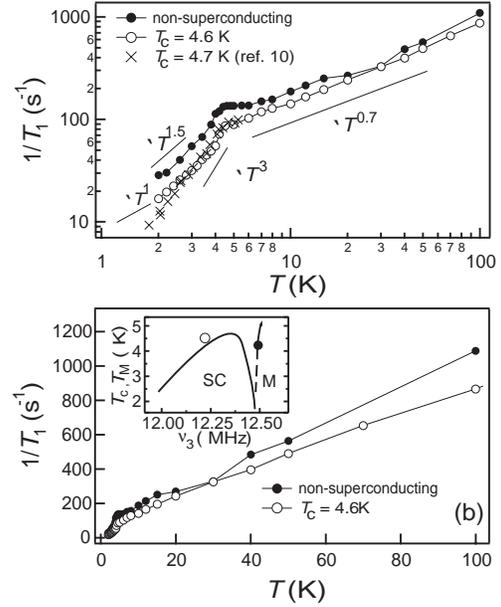}
\end{center}
\caption{\label{fig:invT1}
(a) Logarithmic and (b) linear scales of the temperature dependence of the nuclear spin-lattice relaxation rate (1/$T_1$) in non-superconducting and superconducting BLH with $T\mathrm{_{c}}$ = 4.6 K ($\nu_{3}$ = 12.20 MHz at 10 K, right side in the domed superconducting phase in Fig. \ref{fig:phase}). Inset in Fig. (b) is a sketch of Fig. \ref{fig:phase}. Notations are the same in the main and inset figures. }
\end{figure} 

Figure \ref{fig:invT1} shows the temperature dependence of the nuclear spin-lattice relaxation rate 1/$T_1$ in the non-superconducting and superconducting ($T\mathrm{_{c}}$ = 4.6 K) BLH samples. In both cases, 1/$T_1$ shows a rapid decrease below $T\mathrm{_{c}}$ = 4.6 K and $T\mathrm{_{M}} \sim$ 4.2 K, indicating that the elementary excitation spectrum changes at $T\mathrm{_{c}}$ and $T\mathrm{_{M}}$. At low temperatures in the paramagnetic state, both relaxation rates 1/$T_1$ show non-Korringa power law behavior ($\sim T^{n}$, $n < 1$) and the index $n$ is near or less than 0.7. In the superconducting sample below $T\mathrm{_{c}}$, the power law index $n$ is about 3 without coherence peak, indicating the presence of the line nodes in the superconducting gap parameter. The $T$-linear behavior of $1/T_{1}$ in the wide range of the low temperatures indicates that the residual density of states is lager than that of the optimal superconducting BLH samples previously investigated. \cite{Ishida,Kato} The appreciable residual density of states is a characteristic of the line-node superconductivity. 

In the non-superconducting BLH, $1/T_{1}$ does not show the critical divergence at $T\mathrm{_{M}}$. The $1/T_{1}$ behavior near $T\mathrm{_{M}}$ resembles the case of the charge-density-wave transition. But at least the static internal magnetic field exists below $T\mathrm{_{M}}$. Below $T\mathrm{_{M}}$, $1/T_{1}$ shows the power raw behavior with $n \sim$ 1.5. The $T$-linear behavior is expected in both the itinerant ferromagnetic and itinerant N\'{e}el states in $T \ll T_\mathrm{M}$ according to the self-consistent renormalization theory. \cite{Moriya} Thus, the low lying excitation spectrum of the magnetically ordered state in the non-superconducting BLH is thought to be unconventional. 
\begin{figure}
\begin{center}
\includegraphics[width=0.75\linewidth]{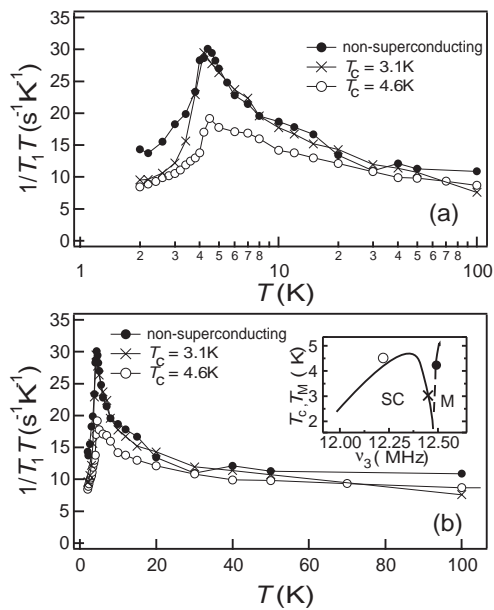}
\end{center}
\caption{\label{fig:invT1T}
(a) Logarithmic and (b) linear scales of the temperature dependence of 1/$T_{1}T$ in non-superconducting and superconducting BLH with $T\mathrm{_{c}}$ = 4.6 K ($\nu_{3}$ = 12.20 MHz at 10 K) and 3.1 K ($\nu_{3}$ = 12.45 MHz at 10 K, right side in the domed superconducting phase in Fig. \ref{fig:phase}). Inset in Fig. (b) is a sketch of Fig. \ref{fig:phase}. Notations are the same in the main and inset figures.}
\end{figure} 

Figures \ref{fig:invT1T} (a) and (b) show the temperature dependence of 1/$T_{1}T$ in logarithmic and linear scales, respectively. In addition to the non-superconducting and superconducting ($T\mathrm{_{c}}$ = 4.6 K) BLH samples in, 1/$T_{1}T$ in the superconducting BLH with $T\mathrm{_{c}}$ = 3.1 K is shown. The sample with $T\mathrm{_{c}}$ = 3.1 K is located near the boundary between superconducting and magnetic phases in the phase diagram in Fig. \ref{fig:phase}. At low temperatures above $T\mathrm{_{c}}$ and $T\mathrm{_{M}}$, 1/$T_{1}T$ increases with decreasing temperature due to the enhancement of the magnetic fluctuations, and decreases at the superconducting and magnetic phase transitions. The enhancement of 1/$T_{1}T$ increases as $\nu_{3}$ increases, suggesting the magnetic correlations develop with $\nu_{3}$. Although $T\mathrm{_{c}}$ increases with the enhancement of 1/$T_{1}T$ and the magnetic correlations in the left side of the optimal $\nu_{3} \sim$ 12.30 MHz in Fig. \ref{fig:phase} as well as the previous report, \cite{Ihara2} the enhancement of 1/$T_{1}T$ becomes larger with increasing $\nu_{3}$ while $T\mathrm{_{c}}$ decreases in the right side of the superconducting phase. We also observed a tendency that the residual density of states (an extrapolation of 1/$T_{1}T$ to 0 K) becomes larger apart from the optimal $\nu_{3} \sim$ 12.30 MHz in Fig. \ref{fig:phase}. 

The hole pocket-like Fermi surface near the $K$ point is thought to be enlarged with the enlargement of the distortion of the CoO$_6$ octahedron, which can be proved by $\nu_{\mathrm{Q}}$. In this case, the magnetic fluctuations near and in $\mathbf{q}$ = 0 are important and may cause the spin-triplet superconductivity. \cite{Mochizuki} The magnetic fluctuations should be necessary for the occurrence of the superconductivity in BLH. However, the superconductivity is suppressed near the phase boundary between the superconducting and magnetic phases, which should be the unconventional magnetic state. It should be stressed that this is not a conventional magnetic quantum critical point. Since the magnetic enhancement of $1/T_{1}$ shown in Fig. \ref{fig:invT1} (b) does not change markedly between the superconducting and magnetic BLHs, not only magnetic correlations but also interlayer correlations related to $\nu_{3}$ may play the significant roles in $T\mathrm{_{c}}$.

We observed both finite signs of the superconducting and magnetic phase transitions in several samples near the boundary between the superconducting and magnetic phases. Because the vacancies of the sodium ions, the contents of the oxonium ions and the water molecules fluctuate structurally in the BLH Na$_{x}\mathrm{CoO_{2}} \cdot$$y\mathrm{H_{2}O}$ system, we could not confirm the coexistence of superconductivity and magnetic ordering in a sample.

In summary, we have done $^{59}$Co NQR studies of BLH sodium cobalt oxides Na$_{x}\mathrm{CoO_{2}} \cdot$$y\mathrm{H_{2}O}$ with the superconducting transition temperatures 2 K $<$ $T\mathrm{_{c}}$ $\le$ 4.6 K as well as the non-superconducting but magnetic BLH. We developed a magnetic phase diagram of $T\mathrm{_{c}}$ and $T\mathrm{_{M}}$ against $\nu_{3}$ which may specify the degree of the distortion of the CoO$_6$ octahedron. We observed a dome shape superconducting phase with the optimal $\nu_{3} \sim$ 12.30 MHz. In the non-superconducting BLH, the $^{59}$Co NQR spectrum shows a broadening below $T\mathrm{_{M}}$ without the critical slowing down effect of 1/$T_{1}$ and 1/$T_{2}$, indicating an unconventional magnetic ordering. From the analysis of the temperature dependence of the relaxation rates, we concluded that the magnetic fluctuations are enhanced above $T\mathrm{_{c}}$ and $T\mathrm{_{M}}$ with the increase of $\nu_3$. We observed that $T\mathrm{_{c}}$ is suppressed near the phase boundary $\nu_3 \sim$ 12.50 MHz, which may arise from the additional effects to cause an unconventional magnetic ordering.

We thank Dr. T. Waki and Dr. M. Kato for their experimental supports, and Dr. K. Ishida, Mr. Y. Ihara and Dr. Sakurai for fruitful discussion.
This work is supported by Grants-in Aid for Scientific Research on Priority Area "Invention of anomalous quantum materials", from the Ministry of Education, Culture, Sports, Science and Technology of Japan (16076210).

\end{document}